\documentclass[doublecol]{epl2}
\usepackage{graphicx}
\usepackage{amsfonts,amsmath,amssymb}
\usepackage[normalem]{ulem}
\usepackage{bbm}

\title{Pairing properties of cold fermions in the honeycomb lattice}

\author{Beno\^it Gr\'emaud\inst{1,2,3}}
\shortauthor{Beno\^it Gr\'emaud}
\institute{
\inst{1}Laboratoire Kastler Brossel, UPMC, ENS, CNRS; 4 Place Jussieu, F-75005 Paris, France\\
\inst{2}Centre for Quantum Technologies, National University of Singapore, 3 Science Drive 2, Singapore 117543, Singapore \\
\inst{3}Department of Physics, National University of Singapore, 2 Science Drive 3, Singapore 117542, Singapore \\
}

\abstract{
The pairing properties of ultracold fermions, with an attractive interaction, loaded in 
a honeycomb (graphene-like) optical lattice are studied in a mean-field approach. 
We emphasize, in the presence of a harmonic trap, the 
unambiguous signatures of the linear dispersion relation of the band structure around half-filling 
(i.e. the massless Dirac fermions) in the local order parameter, in particular in the situations of 
either imbalance hoping parameters or imbalance populations. It can also be observed in the system
response to external perturbation, for instance by measuring the pair destruction rate when modulating the optical lattice depth. 
Going beyond the mean-field level, we estimate the critical temperature for the  "condensation" of the preformed pairs.
}


\begin{document}

\maketitle

Since its first experimental observation, graphene has attracted considerable attention due
to its interest in fundamental physics as well as for potential
applications\cite{Novoselov_04,Neto_09}. In particular, 
the energy band spectrum depicts ``conical points" where the valence and conduction bands are connected, and
the Fermi energy at half-filling is therefore only made of points. Around these points, the
energy varies proportionally to the modulus of the wave-vector and the excitations
(holes or particles) of the system are equivalent to ultra-relativistic (massless) Dirac
fermions since their dispersion relation is linear~\cite{Semenoff_84}. 
In the presence of interaction, the vanishing density of states at 
the Fermi energy leads to extremely rich physics. Contrary to the square
lattice, where the nesting of the Fermi surface generally leads to ordered
phases even for arbitrarily small interaction strengths, the Fermi-Hubbard model on the honeycomb
lattice depicts two quantum phases transitions at half-filling~\cite{Meng_10}: first, for weak interaction, one observes a semi-metallic
behavior with Dirac-like excitations~\cite{Paiva_05,Zhao_06,Lee_09} similar to the non-interacting situation; for an interaction strength $U\approx3.5J$,
the system enters into a spin-liquid phase, i.e. a Mott-Insulator (charge gap) without long-range magnetic ordering~\cite{Hermele_07,Wang_11}; eventually,
for larger interaction strength $U\approx 4.3J$, anti-ferromagnetic order sets in.  

Naturally, these very peculiar properties (massless Dirac fermions, spin-liquid...) have found a large echo in the field of ultracold
atoms in optical potentials~\cite{Zhu_07,keanloon}, which are now widely acknowledged to be great tools to study and
understand the transport or magnetic properties of their condensed matter counterparts~\cite{Dalibard,ketterle2}. This is 
especially true in the case of graphene, since these quantum phase transitions have not been 
observed yet, due to possible discrepancies between graphene and the physics
depicted by the Fermi-Hubbard model and to rather weak electron-electron interaction\cite{Neto_09}; on 
the contrary the interaction strength between cold atoms can be tuned via Feshbach resonance. 
Therefore,  even though atoms in cubic lattices
is still the standard situation, cold gases in the honeycomb lattice start also to be the subject of many experiments,
leading to new phenomena, either for bosons~\cite{Sengstock_11_a,Sengstock_11_b}
or for fermions~\cite{Esslinger_11}. 

Still, ultracold fermions in lattice have few drawbacks: the presence of an external confinement 
and a relatively high temperature~\cite{ketterle2}. Usually, the external confinement is taken into account 
using the local density approximation, i.e. assuming that the local properties 
can be described by the non-confined system ones for the local chemical potential~\cite{Tempere_08,Iskin_07,Iskin_Melo_08}. 
On the other hand, this 
assumption usually breaks down when the fermionic density has strong variation at the lattice scale; this is
typically the situation at the border of the atomic cloud or, in the case of imbalanced population, at the boundaries
between the different phases, like paired-non paired, polarized-non polarized... The proper description of these interfaces
goes beyond the local density approximation. 
In this letter, the properties of a two-component fermionic gas with on-site attraction in the presence of both
a honeycomb lattice and a harmonic trap are discussed in a mean-field approach. To properly account for spatial
inhomogeneities, the BCS order parameter at each site is an independent variable~\cite{Iskin_08,Fujihara_10}, whose value is
determined, for a given temperature, by a global minimization of the free energy. 
After a short description of the method, 
we first describe the behavior of the order parameter in the standard situation: perfect lattice and balanced population; then
we emphasize the specificities of the honeycomb lattice when unbalancing either the hoping parameters or the populations.
Furthermore, we show how response functions of the system can be computed, for
instance, the pair destruction rate induced by a small modulation of the lattice potential. Finally, at
large interaction, we emphasize the different temperature scales, namely the pair formation and their condensation, 
the later transition being of the Berezinsky-Kosterlitz-Thouless (BKT) type, i.e. controlled by the phase fluctuations~\cite{BKT}.

Starting from the full Fermi-Hubbard model on the honeycomb lattice 
(see Ref.~\cite{keanloon} for a detailed description of the geometry), one can derive the mean-field Hamiltonian:
\begin{equation}
\begin{aligned}
 H_{MF} &= \psi^{\dagger}M\psi+\frac{1}{U}\sum_{i}\Delta_i^*\Delta_i-\sum_i\mu_{i\downarrow}\\
M=&\left(\begin{array}{cc}
 h_{ij\uparrow} & -\Delta_i \\
 -\Delta_i^* & -h_{ji\downarrow}	
\end{array}\right),
\end{aligned}
\end{equation}
where $\Delta_i^*=U\langle f^{\dagger}_{i\uparrow}f^{\dagger}_{i\downarrow}\rangle$ are on-site pairing amplitudes;
$\Psi^{\dagger}=\left(\cdots,f_{i\uparrow}^{\dagger},\cdots,f_{i\downarrow},\cdots\right)$ is the Nambu spinor.
The matrix $h$ depicts the one particle Hamiltonian, namely hopping terms between nearest neighbors 
$h_{ij\sigma}=-J_{ij}\quad i\ne j$ and chemical potential terms
$h_{ii\sigma}=-\mu_{i\sigma}=-\mu_{\sigma}+\frac{1}{2}M\omega_{\mathrm{trap}}^2R_i^2$.
In the case of perfect honeycomb lattice, all the tunneling rates $J_{ij}$ have the same value $J$, which
for a typical
value $V_0=10E_R$, is given by $J\approx0.07E_R$, where $E_R=\hbar^2k_L^2/2M$ is the recoil energy~\cite{keanloon}. 
This value of $J$ sets the unit of energy to $210$ nK ($4.2$ kHz) for ${}^6$Li or  $30$ nK ($600$ Hz) for  ${}^{40}$Na.

The free energy $F=-\frac{1}{\beta}\ln{(\mathcal{Z})}$ associated to the mean-field Hamiltonian reads:
\begin{equation}
 F=-\frac{1}{\beta}\sum_k\ln{\left(1+e^{-\beta\lambda_k}\right)}+\frac{1}{U}\sum_{i}\Delta_i^*\Delta_i-\sum_i\mu_{i\downarrow},
\end{equation}
where the $\lambda_k$ are the $2N$ eigenvalues of the Nambu matrix $M$; $N$ is the number of sites
($2\times51\times51$ in the following). The ratio between the harmonic trap energy 
$\hbar\omega_{\mathrm{trap}}$ and the bandwidth $6J$ is $\approx 0.04$, i.e. corresponding to trap frequency
$1.1$ kHz for ${}^6$Li, $160$ Hz for ${}^{40}$Na. 

As explained in the introduction, the values of $\Delta_i$ are obtained by minimizing the free energy using a mixed 
quasi-Newton and conjugate gradient method; additional checks were performed to ensure that the global minimum has been reached.

\begin{figure}
\centerline{\includegraphics[width=6cm]{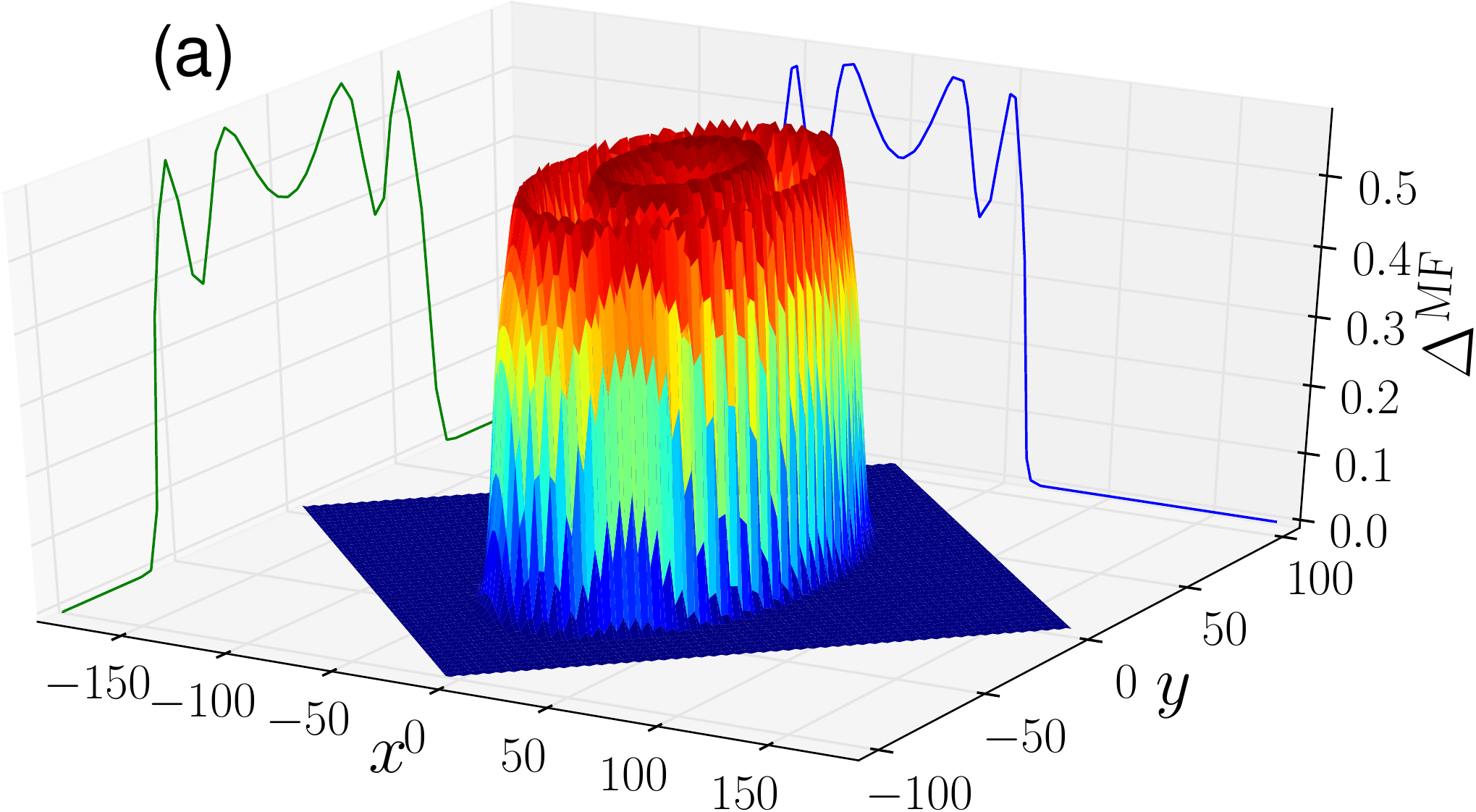}}

\centerline{\includegraphics[width=6cm]{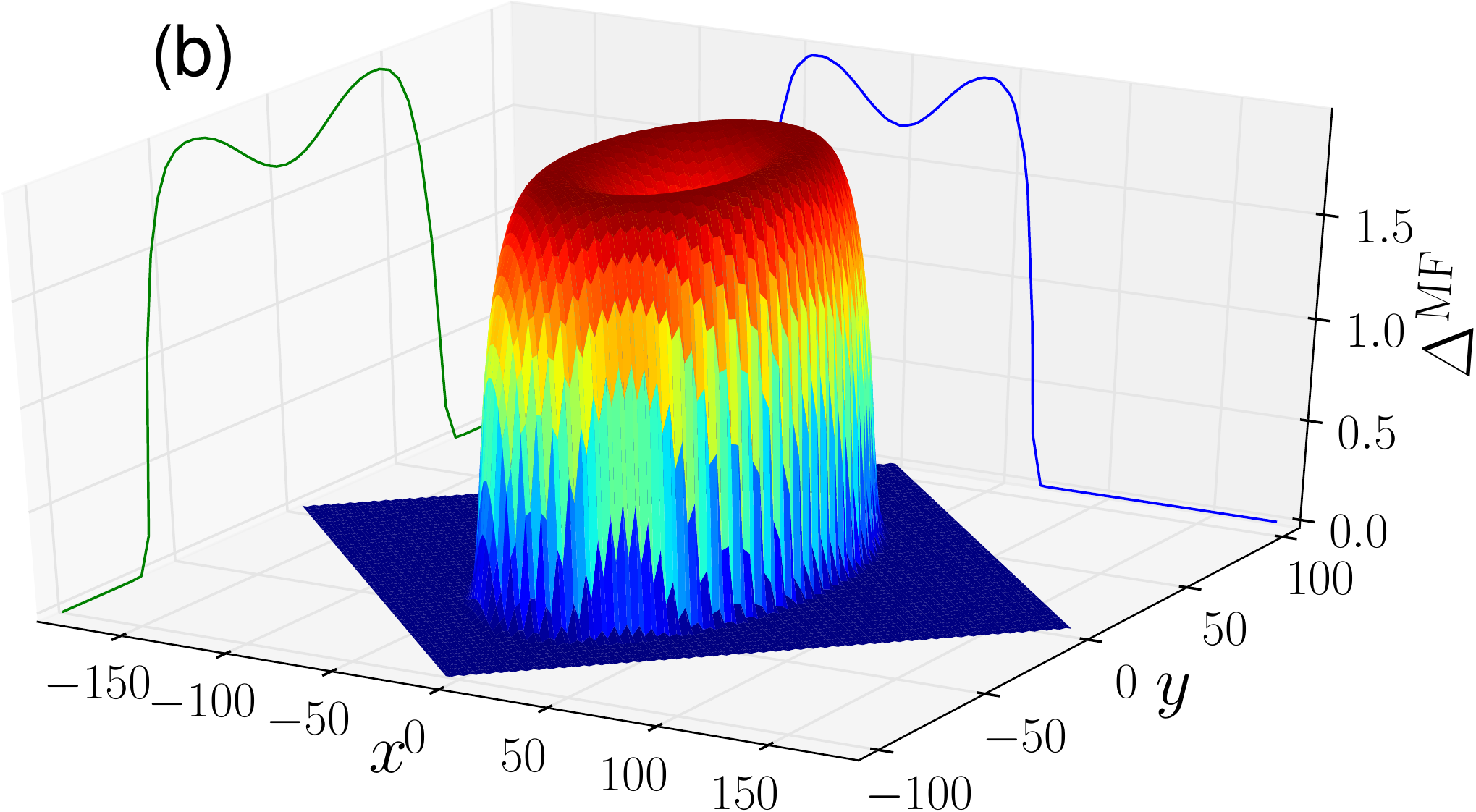}}
\caption{\label{balanced}(Color online) Order parameter distribution in real space for two different values
of the interaction $U=2.5J$ (a) and $U=5J$ (b) ($\beta J=25$). In the first case, the order parameter 
$\Delta$ depicts a dip having a ring-like structure corresponding to a local  half-filling situation,
at which, due to the conical intersections, the pairing is less efficient. At larger interaction, the ring disappear
disappears and the $\Delta$ depicts a smooth pattern. The dip in the center reflects that, in general, the pairing 
efficiency decreases when getting closer to the unit filling.}
\end{figure}

In the situation of balanced spin populations, the distribution of the order parameter is position space is shown in 
fig.~\ref{balanced}, for two interaction values $U=2.5J$ and $U=5J$ ($\beta J=25$). 
The value of the chemical potential at the center of the trap is $\mu=2J$, 
leading to fermionic populations
$N_{\sigma}\approx 677$ (resp. $N_{\sigma}\approx 705$) for $U=2.5J$ (resp. $U=5J$).
For $U=2.5J$ (a), the order parameter $\Delta$ depicts two dips, one in the center and one having a ring-like structure, corresponding exactly 
to a local half-filling situation; this is nothing else but a signature that, at low interaction,
the vanishing density of state at the the conical intersection leads to a poor pairing efficiency.
At larger interaction, $U=5J$ (b), this dip disappears and the $\Delta$ depicts a smooth pattern. 
The dip at the center arises since, for fermions in lattices, the pairing efficiency decreases close to integer filling.
 Therefore, when going from the 
outer part toward the center of the trap, the local number of pairs results from the competition between 
an increase of the number of fermions and a decrease of the pairing efficiency. Nevertheless, for a fixed trap geometry,  
the ratio between the total number of pairs, 
$\sum_i|\Delta_i|^2/U^2$ and the total number of fermion increases with the interaction: $0.07$ for $U=2.5J$,  
$0.15$ for $U=5J$, $0.24 $ for $U=10J$ and $0.3$ for $U=20J$.

\begin{figure}
 \centerline{\includegraphics[width=7cm]{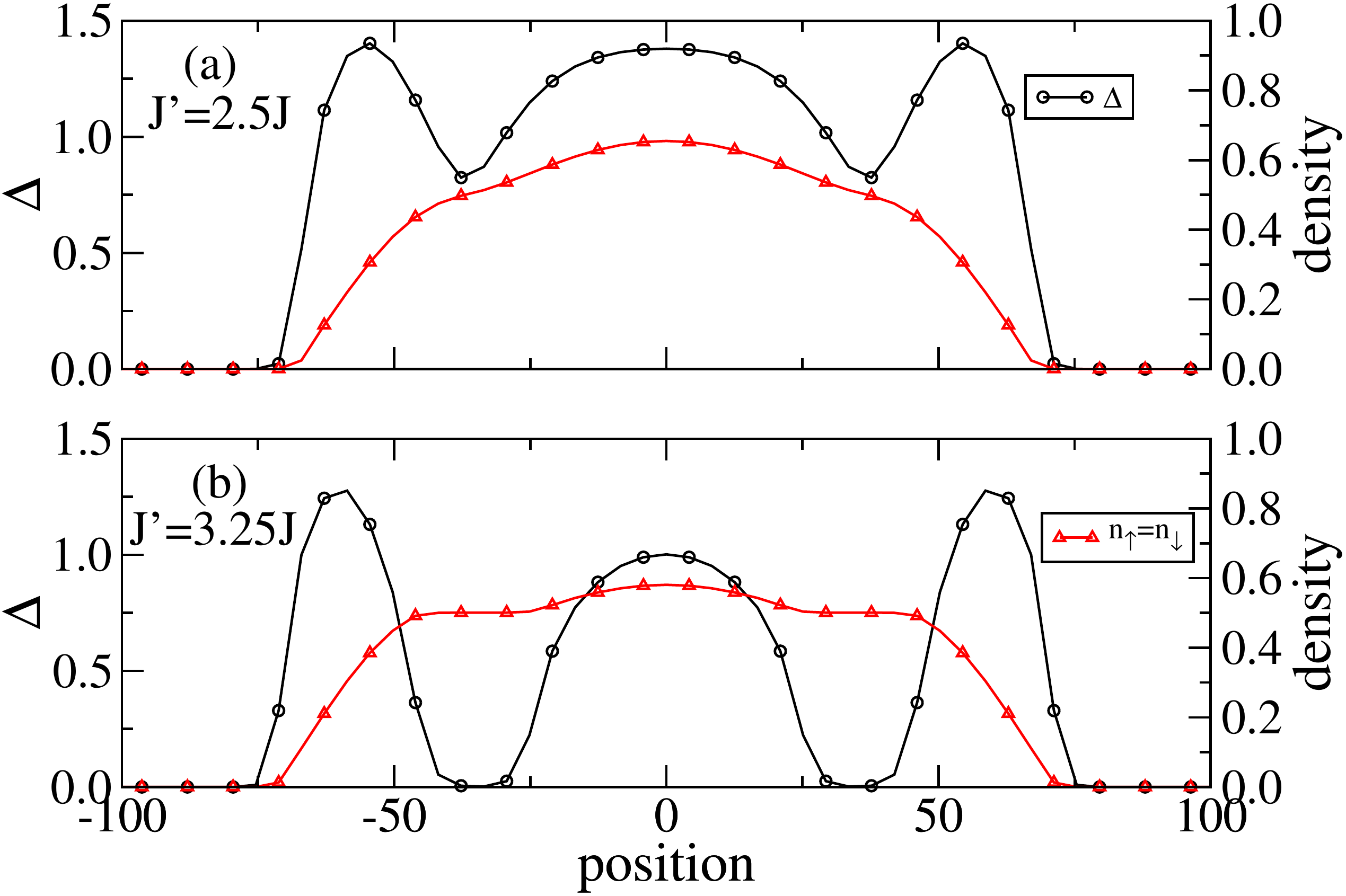}}
\caption{\label{tpot}(Color online) Density and order parameter profiles along a diameter of the trap in the case of
imbalanced hoping parameters ($U=5J$, $\beta J=25$). The black line with the 
circles corresponds to $\Delta_i$ (left axis). The red  line with the  squares
corresponds to fermion density (right axis). (a): a situation where the
gap is smaller than the interaction. The dip in the order parameter takes place around half-filling, where
a kink in the density is already visible. For higher hoping imbalance (b), $\Delta$ now vanishes around half-filling,
the density depicting a clear plateau.}
\end{figure}

\textit{Imbalanced hopings.} In Ref.~\cite{keanloon}, it has been explained that the ratio between the tunneling rates can
be controlled through the lasers building the optical lattice, allowing us to tune the band structure: the conical intersections 
can be moved, eventually merged, opening then a gap, i.e. leading to a two band model. 
This way, one achieves, at half-filling, the transition between the semi-metallic
regime and a band insulator situation~\cite{Montambaux_09,Wang_11}. 
In the presence of an attracting interaction whose strength is smaller than the gap, one expects
to observe  the sequence superfluid-insulator-superfluid when increasing the chemical potential from below to above the band-gap. When the
interaction strength is larger than the gap, the system will always be superfluid, simply the order parameter is expected to become smaller
when the chemical potential lies inside the band-gap. Indeed, in the situation when one of the hoping parameter $J'$ is larger 
than the two others $J$, one observe, for $J'=2.5J$ fig.~\ref{tpot}(a), the presence of a dip
in the order parameter around the position corresponding to half-filling,
where the density depicts a kink, similarly to the non-interactive case~\cite{Zhu_07,Kustblock_10}; 
for higher hoping imbalance, it deepens, eventually leading, around $J'=3.25J$ fig.~\ref{tpot}(b), 
to a vanishing order parameter, i.e., splitting the superfluid in two (independent) components.

\begin{figure}
 \centerline{\includegraphics[width=7cm]{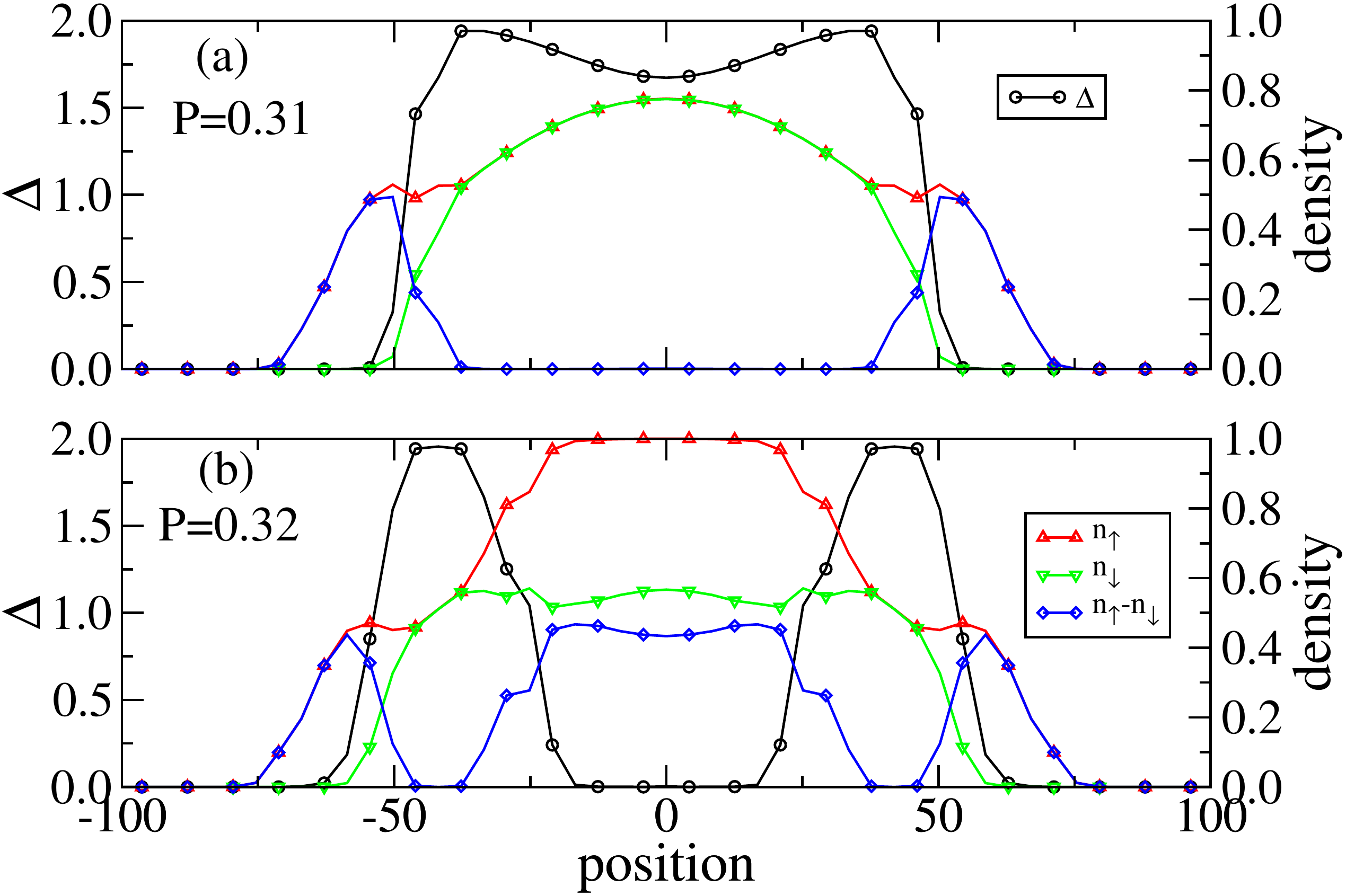}}
\caption{\label{imbalanced}(Color online) Density and order parameter profiles along a diameter of the trap in the case of
imbalanced population ($U=5J$, $\beta J=25$). The black line with the 
circles corresponds to  $\Delta_i$ (left axis). The red  (resp. green) line with upper
(resp. lower) triangles corresponds to  the  up spin (resp. down spin) density (right axis). The blue line with the squares corresponds
to the density difference. (a): polarization $P=0.31$;  
the center of the trap is fully paired and not polarized; the excess fermions are repelled on the edge, 
where one has a fully polarized phase. 
(b): slightly higher polarization ($P=0.32$) the order parameter now depicts a ring-like 
structure; the inner region of the trap is a partially polarized unpaired phase, the majority reaching a unit filling; 
the outer region is a still a fully polarized phase; the intermediate region is roughly made of a fully paired phase.}
\end{figure}

\textit{Imbalanced populations.} The band structure of the honeycomb lattice is peculiar compared to the square lattice one:
(i) there is no Fermi surface nesting at half-filling; (ii) the band structure is not separable, i.e. not given as the sum of
two $1D$ band structures. This results in a quite stable Sarma (breach pairing) phase, without a FFLO-like phase. For instance,
for an interaction strength $U=5J$ and a polarization $P=0.31$, see fig.~\ref{imbalanced}(a), 
one recovers an usual situation~\cite{Zwierlein_science_06,Partrige_2006}: the center of the trap is fully paired and not
polarized, whereas the excess fermions are repelled on the edge, where one has a fully polarized phase. 
For a slightly higher polarization $P=0.32$,  see fig.~\ref{imbalanced}(b), 
there is an abrupt change to the pairing distribution, with the formation of a ring-like
structure of the pairs: the inner region of the trap consists in a partially polarized unpaired phase, the majority reaching a unit filling; 
the outer region is still a fully polarized phase; finally, the intermediate region is roughly made of a fully paired phase, where
the populations of the two fermion components are equal. This in a sharp contrast with the square lattice case, for which one would observe,
for similar parameters, a checkerboard-like pattern for the order parameter, a FFLO-like situation~\cite{Marta_11}.

\textit{Response functions.} At the mean-field level, the Bogolioubov-De Gennes Hamiltonian being quadratic in the Nambu operators $\Psi_{i\sigma}$, 
all response functions, either for imaginary or real frequencies, can be computed from the single particle Green's functions
in imaginary time, $G_{i\sigma\,j\sigma'}(\tau)=-\langle \Psi_{i\sigma}(\tau)\Psi_{j\sigma}^{\dagger}(0)\rangle$, 
where $i$ and $j$ denote two lattice sites and $\langle O\rangle=\mathcal{Z}^{-1}\mathrm{Tr}[Oe^{-\beta H}]$.
Denoting by $P$ the matrix
diagonalizing the Nambu matrix, i.e. $M=P^{\dagger}\Lambda P$, these single particle green's function have the
following expression in imaginary frequency:
\begin{equation}
\label{greenim}
 G_{i\sigma\,j\sigma'}(i\omega_n) = \sum_k \frac{\bar{P}_{k\,i\sigma}P_{k\,j\sigma'}}{i\omega_n-\lambda_k}
\end{equation}
where the $\omega_n=\pi(2n+1)$  denote the fermionic Mastubara's frequencies.

For instance, one accessible experimental quantity is the pair destruction rate when modulating in time 
the depth of the optical lattice, 
which translates into a time dependent modulation of the tunneling 
rate~\cite{Jordens_08,Koroluyk}: $J_{ij}= J^0_{ij}+\delta \cos{(\omega t)}$. 
In the mean-field approach, the local average pair number is 
$|\Delta_i|^2/U^2=\langle f_{i\uparrow}^{\dagger}f_{i\downarrow}^{\dagger}\rangle\langle f_{i\downarrow}f_{i\uparrow}\rangle$,
such that the rate of pair creation/destruction is 
$2\mathrm{Re}(\langle f_{i\uparrow}^{\dagger}f_{i\downarrow}^{\dagger}\rangle\langle \frac{d}{dt}(f_{i\downarrow}f_{i\uparrow})\rangle$.
The later quantity can be derived from
the Heisenberg equations for the local pair operator $f_{i\downarrow}f_{i\uparrow}$:
\begin{equation}
\label{deltarate}
 \langle\frac{d}{dt}(f_{i\downarrow}f_{i\uparrow})\rangle=i\delta\cos{(\omega t)}\sum_{j} J^0_{ij}
\left(\langle f_{i\downarrow}f_{j\uparrow}\rangle+\langle f_{j\downarrow}f_{i\uparrow}\rangle\right)
\end{equation}
the two average values $\langle f_{i\downarrow}f_{j\uparrow}\rangle$ and $\langle f_{j\downarrow}f_{i\uparrow}\rangle$ 
at time $t$ can be obtained from
the linear response theory, assuming that the modulation of the lattice is started at time $t=0$.
\begin{multline}
\label{repfunc}
 \langle f_{i\downarrow}f_{j\uparrow}\rangle(t)=-\frac{\delta}{4\pi}
\biggl[e^{i\omega t}\int d\omega'\frac{e^{i(\omega-\omega')t}-1}{i(\omega-\omega')}\chi_{i\downarrow j\uparrow}(\omega')\biggr.\\
\biggl.-e^{-i\omega t}\int d\omega'\frac{e^{-i(\omega+\omega')t}-1}{i(\omega+\omega')}\chi_{i\downarrow j\uparrow}(-\omega')\biggr]
\end{multline}
where $\chi_{i\downarrow j\uparrow}$, in imaginary time, simply reads:
\begin{multline}
 \chi_{i\downarrow j\uparrow}(\tau)=\sum_{<p,q>\,\sigma} J_{pq}^{0}
\langle f_{i\downarrow}(\tau)f_{j\uparrow}(\tau) f_{p\sigma}^{\dagger}(0)f_{q\sigma}(0)\rangle\\
=\sum_{<p,q>}J_{pq}^{0}
\biggl[G_{p\downarrow i\downarrow}(-\tau)G_{j\uparrow q\downarrow}(\tau)-G_{q\uparrow i\downarrow}(-\tau)G_{j\uparrow p\uparrow}(\tau)\biggr].
\end{multline}
Inserting eq.~\eqref{greenim} in the preceding expression allows us to compute $\chi_{i\downarrow j\uparrow}(\omega)$
as follows:
\begin{multline}
\sum_{kk'\,pq}J_{pq}^{0}
\frac{\bar{P}_{k\,p\downarrow}P_{k\,i\downarrow}\bar{P}_{k'\,j\uparrow}P_{k'\,q\downarrow}}{\omega+i\epsilon+\lambda_k-\lambda_{k'}}
\left(n_f(\lambda_k)-n_f(\lambda_{k'})\right)\\
-\sum_{kk'\,pq}J_{pq}^{0}
\frac{\bar{P}_{k\,q\uparrow}P_{k\,i\downarrow}\bar{P}_{k'\,j\uparrow}P_{k'p\uparrow}}{\omega+i\epsilon+\lambda_k-\lambda_{k'}}
\left(n_f(\lambda_k)-n_f(\lambda_{k'})\right)
\end{multline}
where $n_f(\lambda)$ is the Fermi function at temperature $1/(\beta J)$. 

\begin{figure}
 \centerline{\includegraphics[width=7cm]{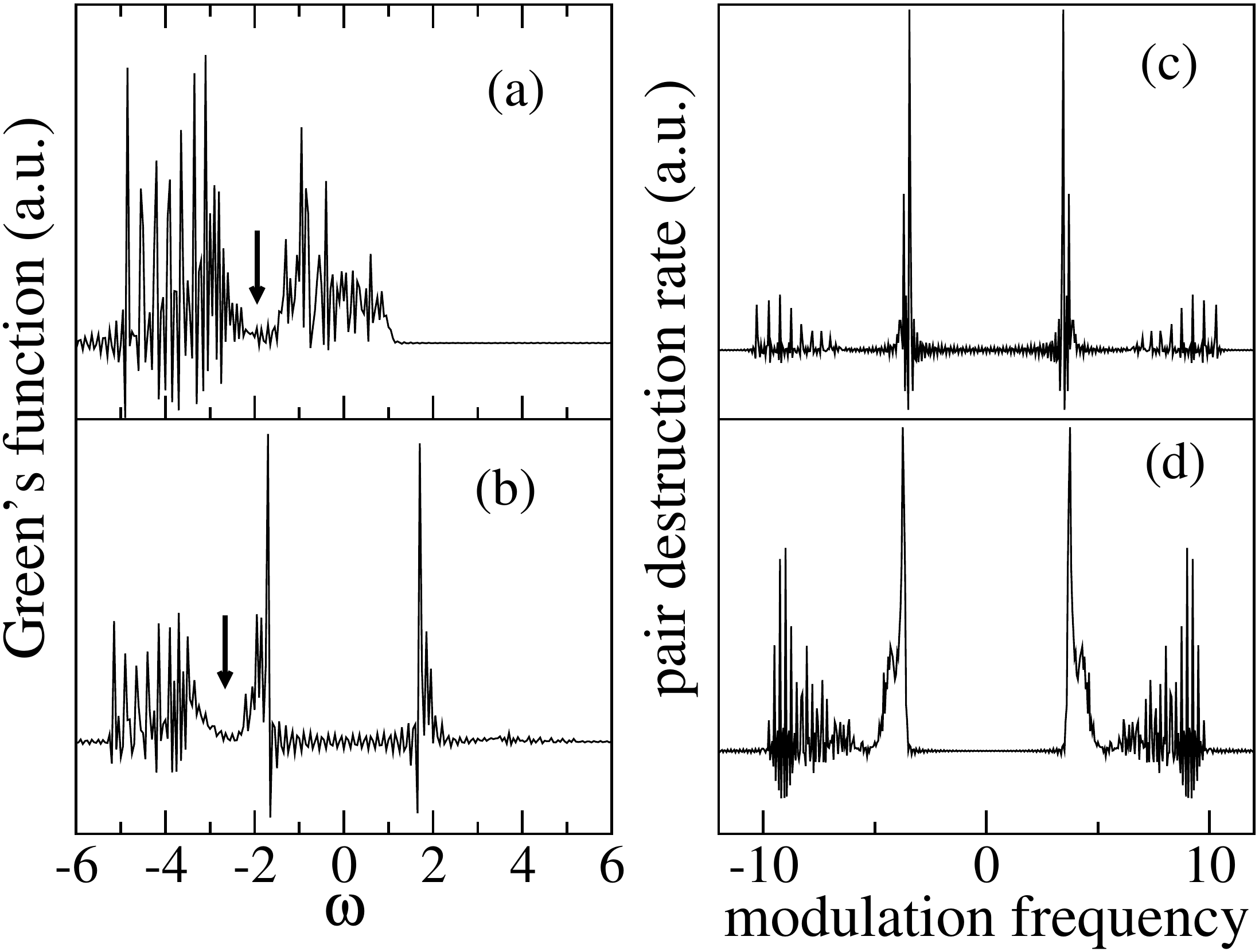}}
\caption{\label{pair_green}(Color online) On the left, the single particle Green's function 
corresponding to a site at the center of the trap are plotted 
for $U=0$ (a) and $U=5J$ (b). For $U=0$, a dip in the excitation spectrum, due to the
conical intersections, appears around $\omega=-2J$. For $U=5J$, this dip moves to
$-\sqrt{\mu_i^2+\Delta_i^2}\approx-2.6J$, because of the BCS gap $\Delta_i$. The peaks structure
corresponds to the harmonic levels of the fermions having an effective mass given by the curvature of the
band around $\mathbf{k}=0$ (see text). On the right, the pair destruction rate when 
modulating the lattice potential is plotted as a function of the modulation frequency. Plot (c)
corresponds to a site a the center of the trap, whereas plot (d) corresponds to a site slightly away from 
the center.  The BCS gap is clearly visible. More interesting is that the peaks due to the harmonic levels and
the dip due to the vanishing density of states around the conical intersections are also visible. }
\end{figure}

From eq.~\eqref{deltarate} and eq.~\eqref{repfunc}, one easily sees that the pair rate, at long time $t$, 
has two components mainly oscillating at $\pm2\omega$ and  one component which is almost time independent. In the limit
$t\rightarrow\infty$, the later is what would be given by the usual Fermi golden rule. Here, we compute the
relevant quantities at a finite time $t=80 \hbar J^{-1}$, larger than the typical timescale of the dynamics, 
but short enough not to resolve the discrete spectrum due to finite size of the system. The resulting pair destruction
rate is displayed in fig.~\ref{pair_green}, right column, for two different site positions, 
whereas the left column depicts the single particle Green's function for a site at the center of the trap,
i.e. $\mathrm{Im}\left(G_{i\sigma\,i\sigma}(\omega)\right)$, for both $U=0$ (a) and $U=5J$ (b), with the same 
frequency resolution as for the pair destruction rates. 
Since at the center of the trap, the chemical potential is $\mu=2J$, excitations, for $U=0$, 
corresponding to the conical intersections are located around $\omega=-2J$, which is indicated by the arrow. 
The peak series at lower frequencies actually corresponds to the levels in the harmonic trap for the fermions having an
effective mass $M^*$ set by the curvature of the band at $\mathbf{k}=0$, namely  $M/M^*=(2\pi)^2/9 \times J/E_R$, leading to an 
effective harmonic trap frequency value 
$\omega^*_{\mathrm{trap}}=\omega_{\mathrm{trap}}\sqrt{M/M^*}=\omega_{\mathrm{trap}}2\pi/3\times\sqrt{J/E_R}$,
which, in the present case, gives rise to effective harmonic levels separated by the energy $0.15 J$.
Since the Green's function shown in the figure corresponds to a site almost at the center of
the trap, only even levels can be excited resulting in a frequency
separation between the peaks equal to twice the effective harmonic trap frequency $\approx 0.3J$. 
Furthermore, one can check that the first peak precisely corresponds to the ground state whose energy is 
$-3J+\hbar\omega^*_{\mathrm{trap}}$, i.e. to a frequency
transition $\omega=-\mu-3J+\hbar\omega^*_{\mathrm{trap}}=-4.85J$. Eventually, the peak structure disappears
 at higher frequencies, where deviations from the quadratic dispersion become more important.
The situation is roughly similar in the interacting case, but
with the addition of a clear BCS gap whose size is $2\Delta_i$. Its presence also shifts the excitations to lower
energies; for instance the rough position of the conical intersection is now given by 
$-\sqrt{\mu_i^2+\Delta_i^2}\approx-2.6J$, as indicated by the vertical arrow. These features also show up on the 
pair excitation rates (fig~\ref{pair_green},right column): whether one considers a site at the center (c) or 
a site away from it (d), one can see the BCS gap, whose size is $4\Delta_i$. For larger frequencies, one can also see the peaks corresponding to the harmonic levels. In
between, the region with low excitation rates corresponds to the vanishing density of state around the the conical intersection; 
this last property is specific to the honeycomb lattice and obviously absent in the case of the square lattice.

\textit{BKT transition.} In the large interaction limit, there are two distinct energy scales:
the pair formation ($k_BT\approx U$) and the pair ``condensation'' ($k_BT\approx t^2/U$), which, in 2D,
is driven by the phase fluctuations of the pair order parameters $\Delta_i$, resulting in a 
 BKT-like transition.

Neglecting quantum fluctuations the partition function reads:
\begin{equation}
 \mathcal{Z}\approx\int\mathcal{D}[\Delta,\Delta^*]e^{-\frac{\beta}{U}\sum_i|\Delta_i|^2}
\mathrm{Tr}\left[e^{-\beta H_{MF}(\Delta,\Delta^*)}\right]
\end{equation}
where $\Delta=(\Delta_1,\cdots,\Delta_N)$. The saddle approximation in this integral leads to the usual BCS gap equations.
In the strong interaction limit, the amplitude fluctuations are gaped, whereas the phase fluctuations are gapless (Goldstone mode).
Writing $\Delta_i=|\Delta_i|e^{i\phi_i}$, one fixes the amplitude $|\Delta_i|$ to its mean-field value and only accounts for the
phase fluctuations~\cite{Mayr_05,Aryanpour_07,Dubi_07}:
\begin{equation}
\label{BKTphase}
 \mathcal{Z}\approx\int d\phi_1,\cdots d\phi_n
\mathrm{Tr}\left[e^{-\beta H_{MF}(\phi_1,\cdots,\phi_n)}\right]
\end{equation}

\begin{figure}
 \centerline{\includegraphics[width=8cm]{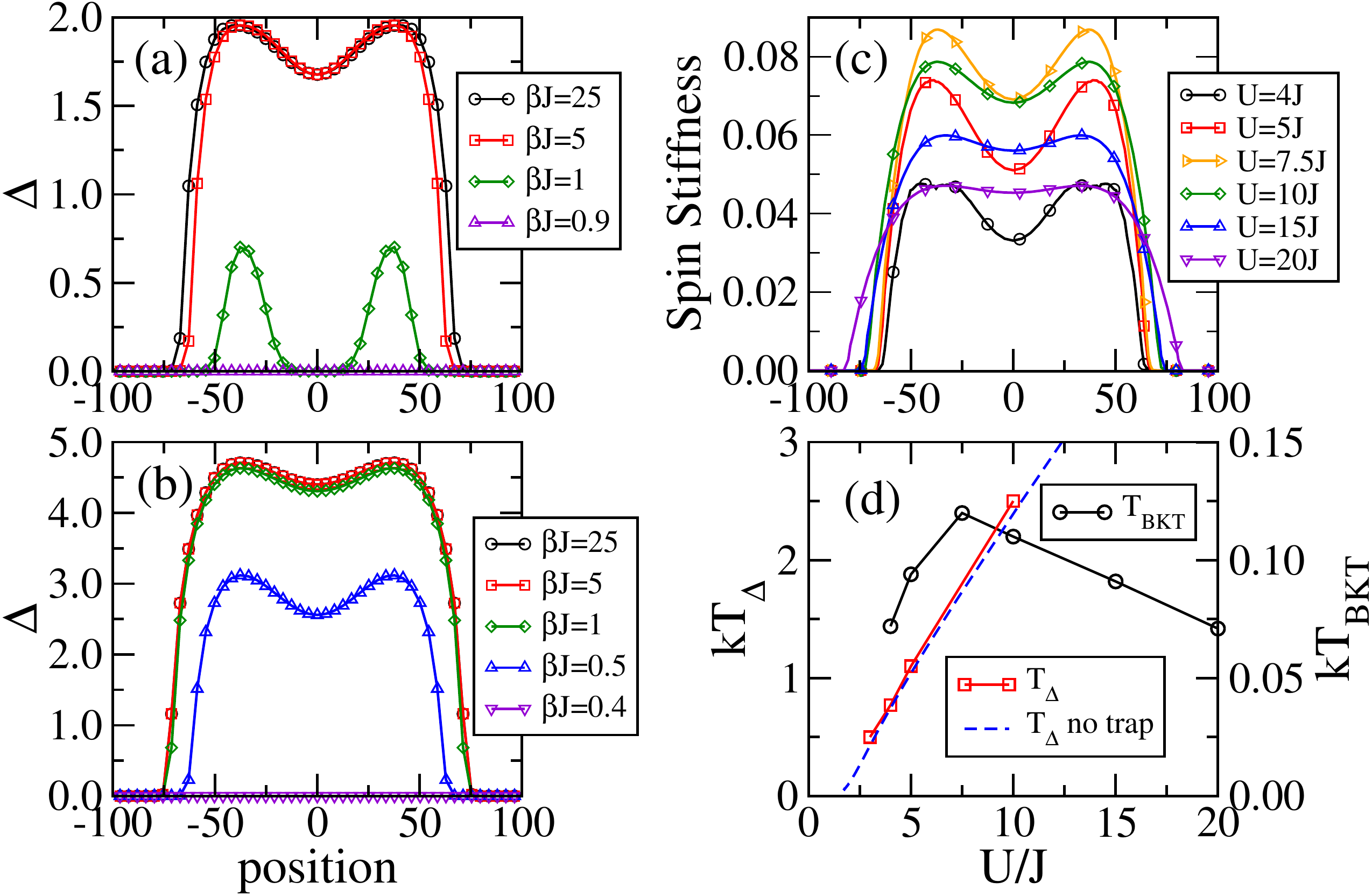}}
\caption{\label{Tcrit}(Color online) The left column depicts the order parameter along a
a diameter of the trap for increasing temperature values (lower $\beta J$ values), 
for two different interaction strengths $U=5J$ (a)
and $U=10J$ (b). For a wide range of temperature,
the profiles are essentially not modified: up to $\beta J=5$ for $U=5J$ and $\beta J=1$ for $U=10J$; 
Furthermore, the order parameter vanishes at higher temperature for larger interaction: 
$\beta J\approx 1$ for $U=5J$ and $\beta J\approx 0.5$ for $U=10J$.
(c): the stiffness parameters $I_{ij}$ for links along a diameter of the trap is plotted
for different values of the interaction $U$. The maximum value is reached around $U=7.5J$, corresponding thus to
the highest temperature of the pair ``condensation''. (d): 
the two temperature scales are plotted as a function of the interaction $U$; 
the ``classical'' pairing $T_{\Delta}$ (red line with squares, left axis) and 
the ``condensation'' of pairs $T_{\mathrm{BKT}}$ (black line with circles, right axis)
The blue dashed line is the ``classical'' pairing critical temperature obtained for the homogeneous system (no trap) 
with the density  corresponding to the one at the center of trap. }
\end{figure}

The BKT temperature can be estimated by expanding the preceding formula at the gaussian 
level\cite{Engelbrecht_97,Iskin_07,Zhao_06,Tempere_08}:
\begin{equation}
\label{spinstiffness}
 \mathcal{Z}\approx\int d\phi_1,\cdots d\phi_n e^{\beta\frac{1}{2}\sum_{ij}I_{ij}\phi_i\phi_j}
\end{equation}
The coefficient $I_{ij}$ describes the local spin stiffness and
 can be evaluated from the mean-field Green's functions~\eqref{greenim}. 
In the homogeneous situation, $I_{ij}=I$, the BKT critical temperature is the simply given by $k_BT_{BKT}=\frac{\pi}{2}I$. 
In fig.~\ref{Tcrit}, the left column shows the order parameter $\Delta$ along a diameter of the trap 
as a function of the inverse temperature $\beta$ for
two different values of the interaction $U=5J$ (a) and $U=10J$ (b). For a wide range of temperature,
the profiles are essentially not modified: up to $\beta J=5$ for $U=5J$ and $\beta J=1$ for $U=10J$; 
In addition, as expected, for larger interaction,
the order parameter vanishes at higher temperature: $\beta J\approx 1$ for $U=5J$ and $\beta J\approx 0.5$ for $U=10J$. 
Figure~\ref{Tcrit}(c)
shows the stiffness parameters $I_{ij}$ for links along a diameter of the trap, 
for different values of the interaction $U$. As expected, it starts from
a low value for $U=4J$, increases up to a maximum value around $U=7.5J$ and then starts decreasing again for larger value.
 Note that for $U=20J$, even though
the profile is different, the value is roughly the same as for $U=4J$. 
Finally, fig.~\ref{Tcrit}(d) summarizes, as a function of the interaction $U$,
the two temperature scales: the ``classical'' pairing $T_{\Delta}$ (red line with squares, left axis) and the ``condensation'' 
(i.e. a quasi-long range order) of pairs $T_{\mathrm{BKT}}$, roughly evaluated from the spin-stiffness.
(black line with circles, right axis). The blue dashed line is the ``classical'' pairing critical temperature 
obtained for the homogeneous system (no trap) having the trap center density.  

\begin{figure}[h]
\centerline{\includegraphics[width=4cm]{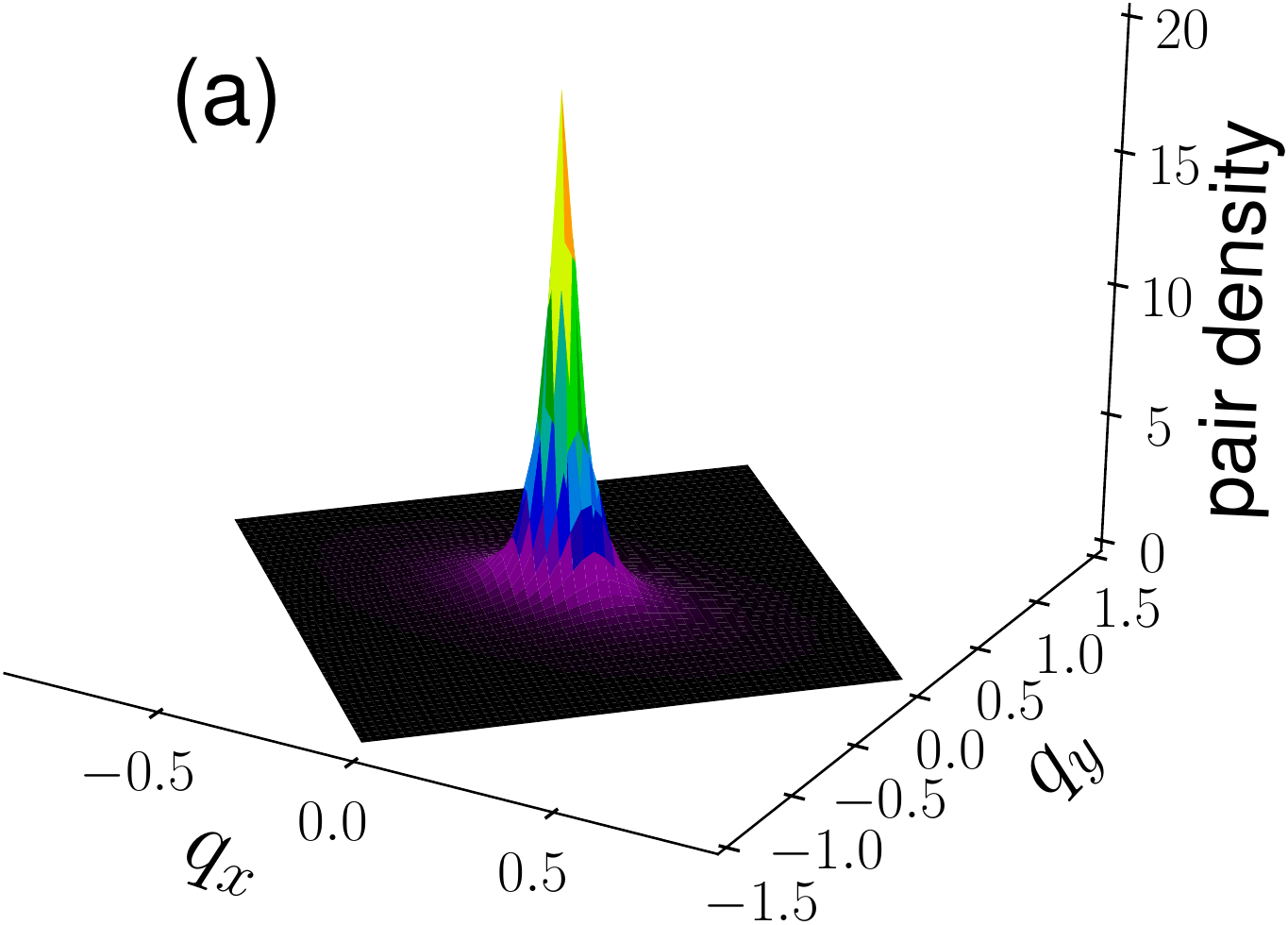}\includegraphics[width=4cm]{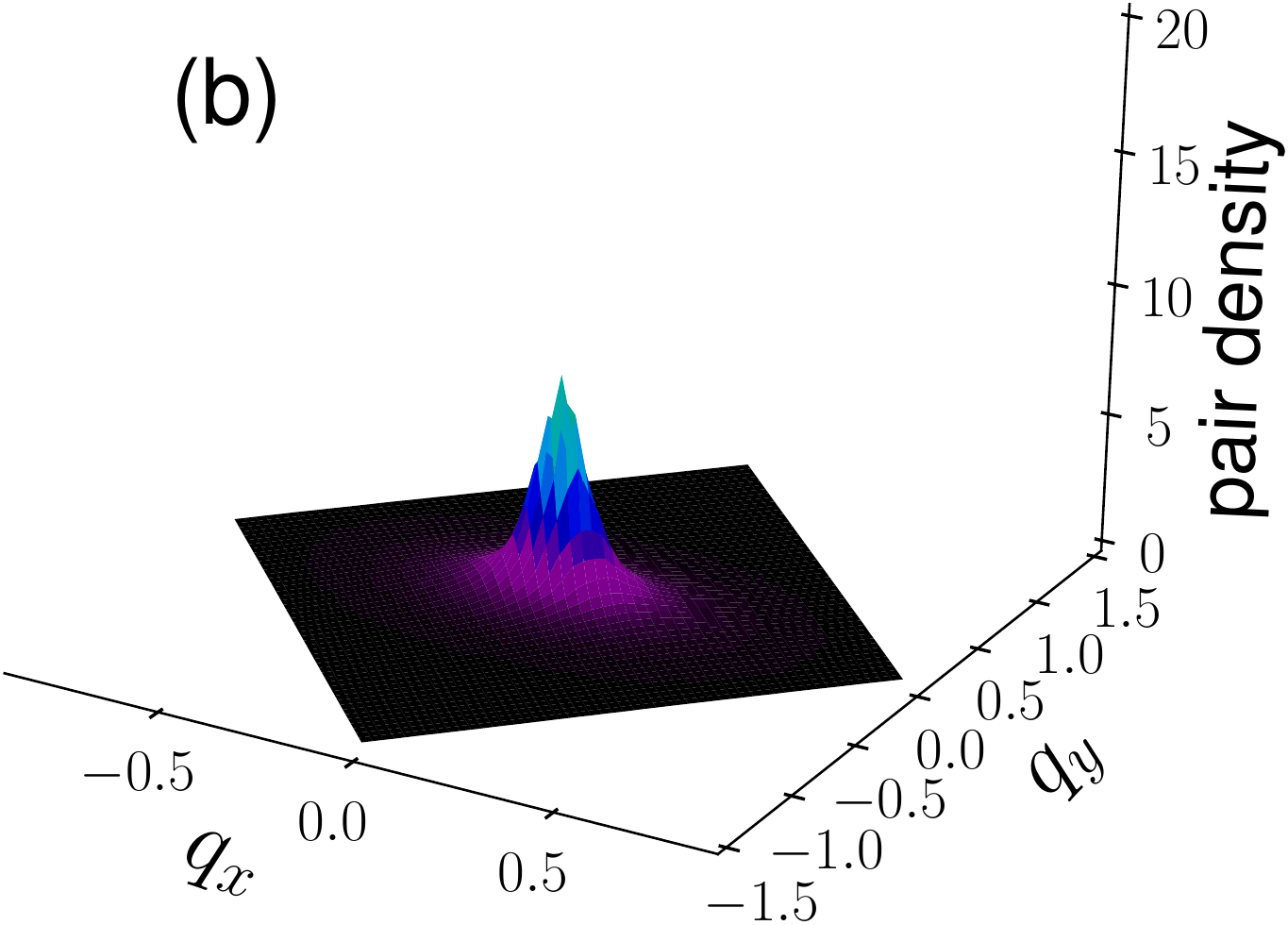}}
\centerline{\includegraphics[width=4cm]{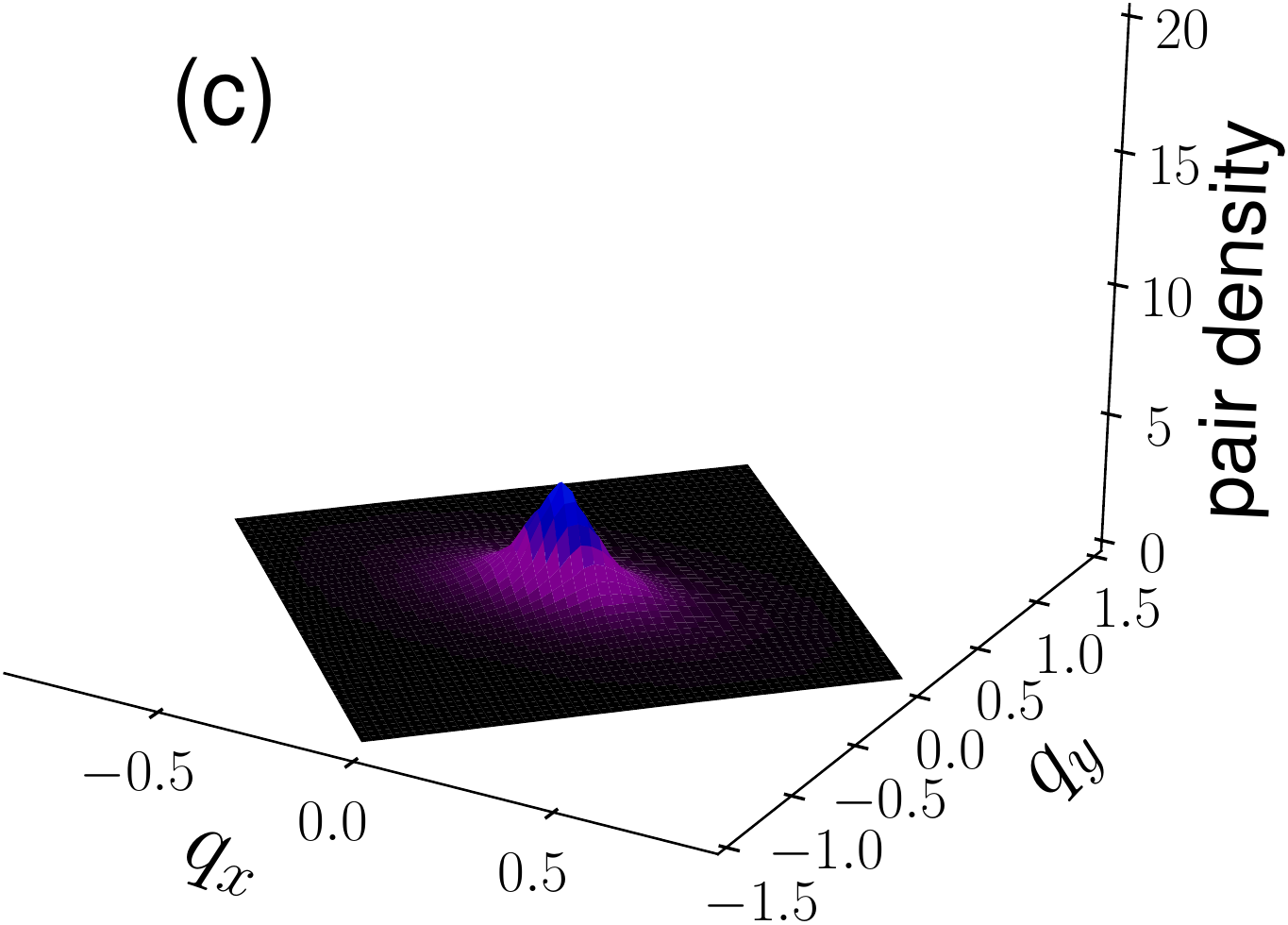}\includegraphics[width=4cm]{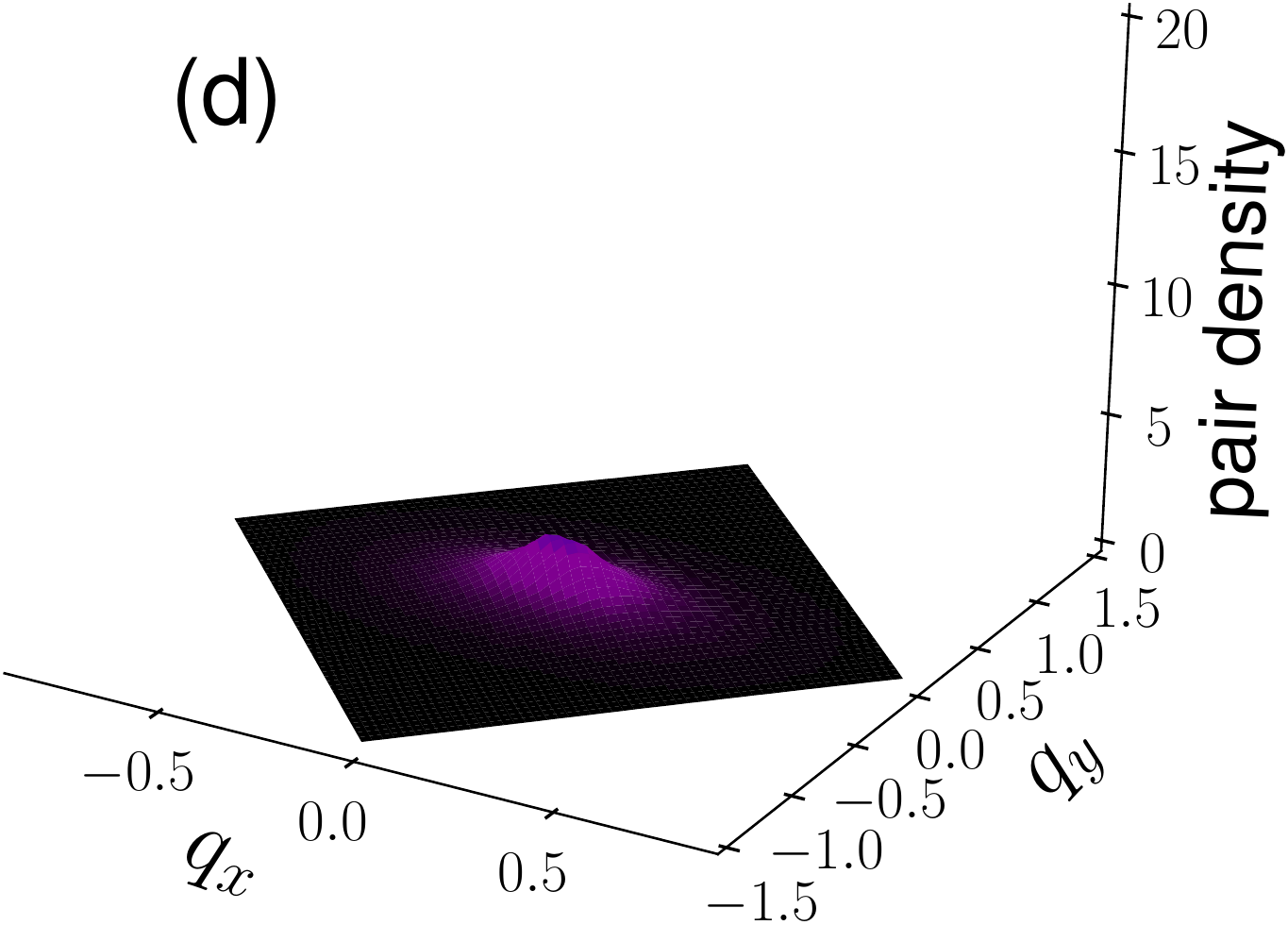}}
\caption{\label{paircondensation} (Color online) Pair density in the quasi-momentum space for different values of the temperature $\beta$,
for $U=5J$;
(a): $\beta J=13$, (b): $\beta J=12$, (c): $\beta J=11$, (d): $\beta J=10$. 
The disappearance of the quasi-condensate peak around $\beta J=11$ is in agreement with the critical value 
$k_BT_{\mathrm{BKT}}\approx0.09J$ extracted from
the typical spin stiffness value, see fig.~\ref{Tcrit}. From the experimental
point of view, this ``condensation'' temperature corresponds to a ratio $T/T_F\approx0.02$.}
\end{figure}

The preceding gaussian approximation does not account for the possibility of vortices in the phases, which are at the 
heart of the BKT transition. In order to cure this problem, one would have to compute the full partition function~\eqref{BKTphase}. 
However, it is numerically too expensive for large lattices. A good approximation consists in using the partition function of the effective XY model 
given by the $I_{ij}$ distribution~\eqref{spinstiffness}. We have checked that for small system, the agreement is fairly good. 
The signature of the transition can be measured in the 
momentum distribution of the pairs. For $T<T_{BKT}$, the quasi-long range order 
$\langle\Delta_i\Delta_j^{\dagger}\rangle\approx |\Delta_i^{MF}||\bar{\Delta}_j^{MF}|\langle e^{i(\phi_i-\phi_j)}\rangle$ leads to a strong peak 
in the momentum distribution of the molecules, i.e. a quasi-``condensate'', see fig.~\ref{paircondensation}. 
This peak broadens and disappear due to the vortex proliferation, for a value of the temperature $\beta J\approx11$ ($U=5J$), in agreement with the 
critical value $k_BT_{\mathrm{BKT}}\approx0.09J$ extracted from the spin stiffness, see fig.~\ref{Tcrit}. From the experimental
point of view, this ``condensation'' temperature corresponds to a ratio $T/T_F\approx0.02$, 
a value $T/T_F\approx0.2$ is enough to the observe the pairing formation, thus, within experimental reach.

In conclusion, we have presented pairing properties of ultracold fermions in a honeycomb lattice and in a harmonic trap with on-site
attractive interaction. We have shown
that one of the main  feature of the honeycomb lattice, namely the conical intersection, leads to unambiguous signatures
in the pair distribution. 
For instance, at low interaction, a circular dip in the pairing distribution appears around the 
local half-filled situation. This dip can even be deepened by
unbalancing the hoping parameters, eventually leading to a two component superfluid.
In the case of imbalanced populations, the absence of Fermi surface nesting gives rise to 
ring-like structure of the pairing, rather than a checkerboard pattern (FFLO situation). 
Furthermore, this vanishing density of state can also be observed in the spectrum of response functions like
the pair creation rate when modulating the optical lattice depth. Finally, we have estimated  the two type of
transition temperatures, namely the pair formation and the pair condensation, emphasizing that for actual experiments
the observation of the paring is definitely within reach, whereas reaching the pair condensation regime still require
a slight improvement in lowering the temperature. Among the possible extensions of the present work, one can mention the
study of fermions in effective gauge fields~\cite{Gorecka_11} or the study of localization properties 
in the presence of disorder\cite{Neto_09}. In addition, genuine dynamics can be studied
in a time-dependent Bogolioubov-De Gennes approach~\cite{Challis_07}.

Center for Quantum Technologies is a Research Centre of Excellence funded by 
the Ministry of Education and National Research Foundation of Singapore.
 This work has been supported by the LIA FSQL.
The author thanks C.A.R.~S\'a~de~Melo, C.~Miniatura and K.L.~Lee for useful discussions.

\end{document}